\documentclass[aps,twocolumn,preprintnumbers]{revtex4}
\usepackage{amsmath}
\usepackage{amssymb}
\usepackage{bbold}
\usepackage{mathrsfs}
\usepackage{graphicx, psfrag}
\usepackage[centerlast]{caption}
\usepackage{float}
\usepackage{subfig}
\usepackage{tikz}
\usepackage{pgfplots}
\usepackage[colorlinks=true, citecolor=blue, urlcolor = blue, linkcolor= red, bookmarks=true]{hyperref}
\usepackage{epstopdf}
\begin{document}
\newcommand{\ts}{\textsuperscript}
\def \beq{\begin{equation}}
\def \eeq{\end{equation}}
\def \bse{\begin{subequations}}
\def \ese{\end{subequations}}
\def \bea{\begin{eqnarray}}
\def \eea{\end{eqnarray}}
\def \bs{\boldsymbol}
\def \bb{\bibitem}
\def \nn{\nonumber}

\title{\textbf{Quantum transport through pairs of edge states of opposite chirality at electric and magnetic boundaries}}
\author{Puja Mondal$^{1}$, Alain Nogaret$^{2}$ and Sankalpa Ghosh$^{1}$}
\affiliation{ $^{1}$ Department of Physics, Indian Institute of Technology Delhi, New Delhi-110016, India}
\affiliation{$^{2}$Department of Physics, University of Bath, Bath BA2 7AY, UK}
\begin{abstract}
We theoretically investigate electrical transport in a quantum Hall system hosting bulk and edge current carrying states. Spatially varying magnetic and electric confinement creates pairs of current carrying lines that drift in the same or opposite directions depending on whether confinement is applied by a magnetic split gate or a magnetic strip gate. We study the electronic structure through calculations of the local density of states and conductivity of the channel as a function of the chirality and wave-function overlap of these states.  We demonstrate a shift of the conductivity peaks to high or low magnetic field depending on chirality of pairs of edge states and the effect of chirality on backscattering amplitude associated with collisional processes. 
\end{abstract}
\maketitle
\section{Introduction}
In a spatially modulated transverse magnetic field, electrons acquire guiding center drift velocity due to the magnetic field gradient even in the absence of an electric field \cite{Muller, Reijniers, Schuler, Park, Sim}. This strongly modifies quantum transport of two-dimensional electron gas (2DEG) due to the formation of chiral current carrying states in the otherwise insulating bulk in the quantum Hall system \cite{Nogaret-c}.  However, compared to the prototype quantum Hall system in uniform magnetic field \cite{Halperin, Macdonald}, magnetically modulated quantum Hall system has received much less attention. There have been interesting experimental development through the observation of asymmetric magnetoconductance peak in a quantum wire \cite{Tarasov}, resistance resonance effect due to magnetic edge states \cite{Nogaret-a, Lawton}, magnetoresistance oscillations as a result of commensurability effect \cite{Weiss, Carmona-a, Ye, Zhang}, giant magnetoresistance \cite{Nogaret-d, Overend, Papp, Xiao,Yoo} and transport assisted by snake orbit \cite{Hara,Schluck, Leuschner, Hoodboy}. The properties of graphene in a non-uniform magnetic field have also attracted attention \cite{Egger,Hausler,Tahir, Masir, Shen, Vasilopoulos, Martino, Huo,Sandner, Yagi, Thomsen, Agrawal}.  In recent work, magnetoresistance anomaly has been observed in a quantum Hall system due to the controlled interference between magnetic edge states and conventional electrostatic edge states \cite{Nogaret-b}.  

In this paper we theoretically investigate two representative hybrid ferromagnetic-semiconductor structures in which magnetic and electrostatic edge states propagate parallel or antiparallel to one another.  These edge states propagate in the same direction under a magnetic split gate whereas they propagate in opposite directions under a bar magnet.  Using an electrostatic gate to gradually deplete the sample edges, we are able to control the overlap between the wavefunctions of electrostatic and magnetic edge states.  This allows us to theoretically study the effect of chirality on the collisional conductivity in the regime of quantum transport. We find that when edge states are far apart quantum transport is adiabatic (the full suppression of inter-edge channel scattering \cite{Beenakker}).  The amplitude of magnetoresistance oscillations is independent of the magnetic potential.  As edge state overlap increases, the edge states which drift in opposite directions give magnetoresistance resistance oscillations which rapidly increase in amplitude.  In contrast no change is observed when edge states drift in the same direction. 
 This leads us to conclude that backscattering between counter-propagating edge states enhances the collisional conductivity of the strip gate device whereas elastic scattering between edge states drifting forward show little change in conductivity in the magnetic split gate.  Hence we predict that transport measurements can evidence the chiral/non-chiral nature of one-dimensional localized edge states.  Our calculations are based on energy levels and wave functions calculated in realistic magnetic potentials.  The density of states (DOS) and local density of states (LDOS) shows the formation of magnetic minibands and electrostatic bands in both magnetic split and strip gate systems.  These calculations demonstrate the formation of a subset of magnetic minibands which arise from quantum interferences between magnetic edge states.
 
Accordingly, the rest of the paper is arranged as follows. In section II, we model the electronic structure of the magnetic split and strip gate and the methodology to solve the Hamiltonian. The local density of states and density of states was calculated to compare the electronic structure of both devices. We discuss the effects of lateral confinement on the LDOS and DOS. In section III, we compute the conductivity tensor within the quantum Boltzmann equation and model the effect of edge state chirality on the disorder conductivity. We describe the resistivity in both magnetic split and strip gate for the decreasing values of lateral confinement.
\section{Electronic structure of the split and strip gates}
\subsection{Model Hamiltonian} 
\begin{figure}
\subfloat{\includegraphics[width=1\columnwidth,height=1
\columnwidth]{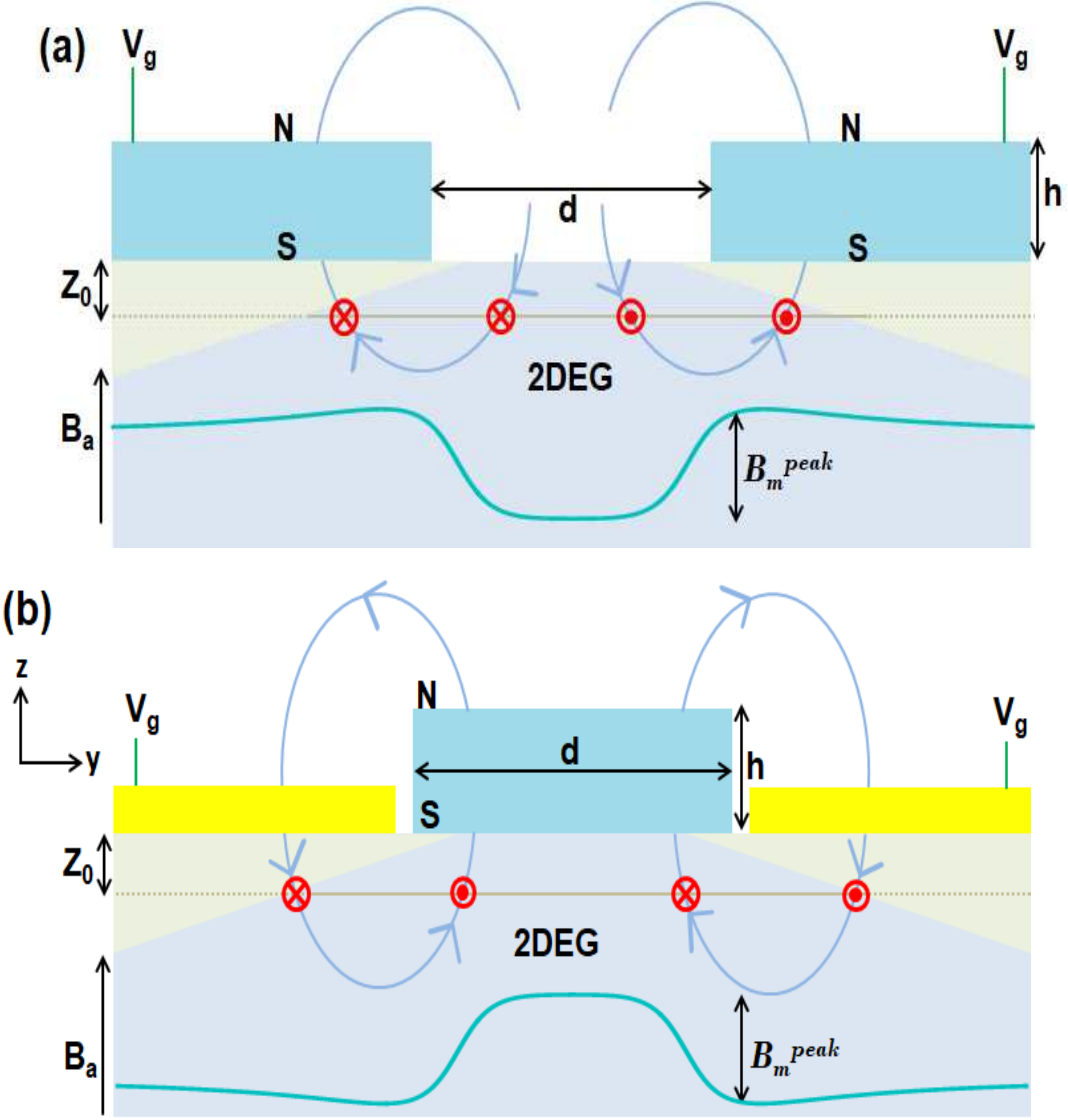}}
\caption{\textit{(color online)} Schematic diagram of the quantum Hall devices (a) magnetic split gate and (b) magnetic strip gate. The stray magnetic field originating from magnetic gate modulates the 2DEG.  Green and yellow slabs represent the ferromagnet and normal metal gate. The approximate position and drift direction of edge states is shown by the arrows (red color). The electrical width of the 2D channel $W_e$ is controlled by the gate voltage $V_g$.  The depletion region underneath the gate is shown by the light brown line. $ B_{m}^{peak} $ is the peak value of the modulated magnetic field.}
\label{fig:schematic}
\end{figure}
\begin{figure}
\subfloat{\includegraphics[width=1\columnwidth,height=1.2
\columnwidth]{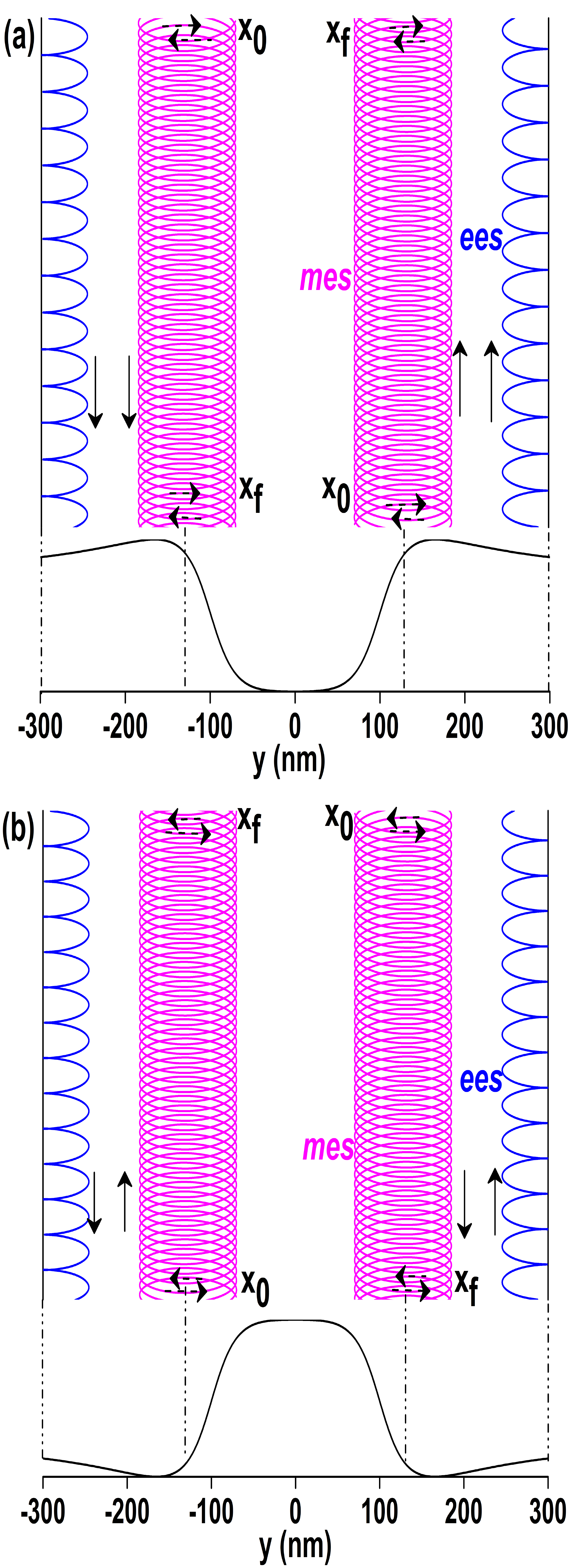}}
\caption{\textit{(color online)} Semiclassical picture of edge states in the magnetic split (a) and strip (b) gate for $\mu_{0}M_{s}$=2.90 T and $B_{a}$=2 T. {\it mes} and {\it ees} are the classical orbit of magnetic and electrostatic edge states. Modulation magnetic field is shown by the black curve.}
\label{fig:classical_orbit}
\end{figure}

 
We consider a 2DEG modulated by a perpendicular magnetic field in two different ways as depicted in Fig. \ref{fig:schematic}.  By using a magnetic split gate or a magnetic strip gate, two magnetic modulations can be produced which have inverted profiles.  A bias voltage is applied to the split magnetic gate in Fig. \ref{fig:schematic}(a) and to the normal metal gate sandwiching the ferromagnetic gate in Fig. \ref{fig:schematic}(b) to deplete the 2DEG underneath.  Through the combination of electrostatic and magnetic potentials, both systems confine electrostatic and magnetic edge states (red arrows).  In the magnetic split gate (a), electrostatic and magnetic edge states always drift in the same direction.  In the magnetic strip gate (b), inverted magnetic field gradient causes the magnetic edge states to drift in the opposite direction to the electrostatic edge states.  Therefore this system is important to study both edge vs bulk conduction and conduction via chiral vs non chiral pairs of edge states.
We write the total magnetic field as:
\beq
 B(y)=B^{\upharpoonleft \! \upharpoonright/\downharpoonright \! \upharpoonright}_{m}(y)+B_{a}
\label{magneticfield}
\eeq
where $B^{\upharpoonleft \! \upharpoonright/ \downharpoonright \! \upharpoonright}_{m}$ and $ B_{a}$ are the modulated and uniform magnetic field. $ \upharpoonleft \! \upharpoonright/\downharpoonright\! \upharpoonright $ refers to the co-propagating (split gate) or counterpropagating edge states (strip gate) as plotted in Fig. \ref{fig:classical_orbit}. 
\bea
B^{\upharpoonleft \! \upharpoonright}_{m}(y)& =& -\frac{\mu_{0}M_{s}}{2\pi}[f_{0}^{+}(y)-f_{0}^{-}(y)-f_{h}^{+}(y)+f_{h}^{-}(y)]\nn\\
B^{\downharpoonright\! \upharpoonright}_{m}(y)& =& \frac{\mu_{0}M_{s}}{2\pi}[f_{0}^{+}(y)-f_{0}^{-}(y)-f_{h}^{+}(y)+f_{h}^{-}(y)]\nn\\
\label{spst} 
\eea
where ${\it f_{z=\lbrace 0,h \rbrace}^{\pm}(y)=atan\bigg{(}\frac{y\pm d/2}{z_{0}+z} \bigg{)}}$. We define {\it h} as the thickness of the magnet, {\it d} is the distance between the two magnets (split gate)or the width of the magnet (strip gate), ${\it z_{0}}$ is the depth of the 2DEG and ${\it \mu_{0}M_{s}} $ is the saturation magnetization.  As a representative case, we choose some experimentally realizable values as {\it d} =200 nm, {\it h}=80 nm, ${\it z_{0}}$=30 nm and ${\it \mu_{0}M_{s}} $=2.90 T (Dy) \cite{Nogaret-b}. The peak value of the modulated magnetic field (green curve in Fig. \ref{fig:schematic} (a) and (b)) is estimated from Eq. \ref{spst} and found $ B_{m}^{peak} $= 0.65 T for both devices as can be obtained from dysprosium magnets. 

The modulated magnetic field changes sign twice while varying over the width of the Hall channel. The distance between lines of zero magnetic field where the magnetic modulation changes sign is defined as $W_{m}$. From the modulated magnetic field profile (Fig. \ref{fig:classical_orbit}), $W_{m}$ can be calculated as 274 nm.  This system has finite magnetic field gradient centered at $ y \approx \pm W_{m}/2$. The gate voltage induces an electrostatic potential to the system resulting in the depletion of 2DEG. We model the electrostatic potential as a square well potential of width $W_e$ :  
\beq
V(y) = \left\{\begin{array}{lr}
        V_{0}=700 meV , & \text{for }\vert y\vert\geqslant W_{e}  \\
         0,  & \text{for} \vert y\vert\leqslant W_{e}
        \end{array} \right\}
        \label{eq:scaler_pot}
\eeq
The square well modelling of electrostatic potential is a good approximation of the transverse potential as the variation of depletion potential at the edge varies fast on the scale of $W_e$. We are going to change $W_e$ to vary the electrical width of the 2DEG by biassing the gate negatively relative to the current contact on the 2DEG. We choose few representative values of $ W_{e} $ from the recent experiment as 600nm, 400nm and 250nm.

The Hamiltonian of a two dimensional electron gas (2DEG) in such magnetic modulation and electrostatic confinement is
\beq
H=\frac{1}{2m^{\ast}}\bigg[p_{y}^{2}+(p_{x}+eA_{x}(y))^{2}+V(y)\bigg]
\label{equ:hamiltonian}
\eeq
where $A_{x}(y)$ is the vector potential corresponding to the magnetic field  given in Eq. \ref{magneticfield}, $m^{\ast}$ is the effective mass of electron. The vector potential is written in the Landau gauge, namely ${\bf A}=(A_{x}(y), 0, 0)$ with  
\beq
A_{x}(y)=A_{x}^{0}(y)+A_{x}^{M}(y)
\eeq
where $A_{x}^{0}(y)$ and $A_{x}^{M}(y)$ are the vector potentials of the uniform and modulated part of the magnetic field. In the Landau gauge, one can write the wave function in the form $\Psi(x,y)=e^{ik_{x}x}\psi(y)$ as the particle has a free motion in the x direction.
We obtain $\psi(y)$ and the corresponding energy levels by numerically solving the Schr\"{o}dinger equation.  Our numerical calculation was done through the relaxation method \cite{Press}. This was necessary to model accurately the effects of the magnetic and electrostatic potential which here are of the same order of magnitude, a problem not suitable for a perturbation treatment.
\beq
\bigg{[}\frac{\partial^{2}}{\partial \bar{y}^{2}}-(\bar{k}_{x}+\bar{A}_{x}(\bar{y}))^{2}-\bar{V}(y)+\bar{E}_{n}(k_{x})\bigg{]}\psi_{n}(\bar{y})=0
\label{Sch-1}
\eeq
where $\bar{k}_{x}=k_{x}l_{b}$, $\bar{y}=y/l_{b}$, $\bar{A}_{x}(\bar{y})=A_{x}(\bar{y})/B_{0}$, $\bar{E}_{n}(k_{x})=\frac{E_{n}(k_{x})}{\hbar^{2}/(2m^{\ast}l_{b}^{2})}$, $\bar{V}(y)=\frac{V(y)}{(\hbar^{2}/2m^{\ast}l_{b}^{2})}$. We define $l_{b}=\sqrt{\hbar/eB_{0}}$=25.66 nm for an uniform magnetic field of strength $B_{0}$=1 T and all the other magnetic fields are measured in units of $B_{0}$. Length and momentum are expressed in units of $l_{b}$ and $l_{b}^{-1}$. The unit of energy is $E_{0}= \frac{\hbar^{2}}{2m^{\ast}l_{b}^{2}}=0.8622~\text{meV}$.  The wave-vector dependent effective potential is of the form $ V_{eff} (\bar{y},\bar{k}_{x})=(\bar{k}_{x}+\bar{A}(\bar{y}))^{2} $ which has a reflection symmetry when changing the sign of $k_x/y $ giving energy spectrum symmetric about the center of the Brillouin zone.  
  
Modulated magnetic field support edge states at the center of the wire drifting in the positive and/or negative x-direction depending on the sign of magnetic field gradient at y=$W_{m}/2$. The drift velocity of semiclassical orbits at the Fermi level in the magnetic field gradient is \cite{Jackson} 
\bea
v_d &=&  \frac{\omega_{0} r_{g}^{2}}{2}\frac{ \nabla \bs B_{m}  \times \bs B_{m} }{(B_{m})^{2}}\nn
\label{drift_velocity}
\eea 
Where $B_m$ is the modulated magnetic field generated by the magnetic gate, $ \omega_{0} $ and $ r_{g} $ are the gyration frequency and radius. The magnetic field gradient ($\nabla B_{m}$) at y=$W_{m}/2$ is positive for the magnetic split gate (Fig. \ref{fig:classical_orbit} (a)). 
Therefore, magnetic edge states drift in the positive x-direction causing parallel motion of the magnetic edge states respecting to electrostatic edge states. While, in the magnetic strip gate, negative magnetic field gradient at y=$W_{m}/2$ results in the anti parallel motion of magnetic edge states relating to electrostatic edge states (Fig. \ref{fig:classical_orbit} (b)).     
\subsection{Electronic band structure}
The formation of edge states at the center of the quantum Hall system is demonstrated through the calculated local density of states (LDOS) \cite{Miller, Fu}. LDOS is obtained from the energy and the eigenfunction of the 2DEG and is defined as 
\beq
\rho(E,y)=\sum_{\alpha}\delta(E-E_{\alpha})\vert \psi_{\alpha}(y)\vert^{2}
\label{LDOS-1}
\eeq    
where $\alpha=\lbrace n,k_{x} \rbrace$ is a quantum state. Impurities introduce collisional broadening of discrete energy levels which we model as:
\beq
\rho(E,y)=\sum_{n,k_{x}}P_{imp}(E-E_{n,k_{x}})\vert \psi_{n}(y,k_{x})\vert^{2}
\eeq
where $P_{imp}(E-E_{n,k_{x}})=\frac{1}{\Gamma \sqrt{\pi}}\exp\bigg(-\frac{(E-E_{n,k_{x}})^{2}}{\Gamma^{2}}\bigg)$ is the Gaussian broadening induced by the impurity with $\Gamma$ being full width at half maxima. 
\begin{figure*}
\subfloat{\includegraphics[width=2\columnwidth,height=0.6
\columnwidth]{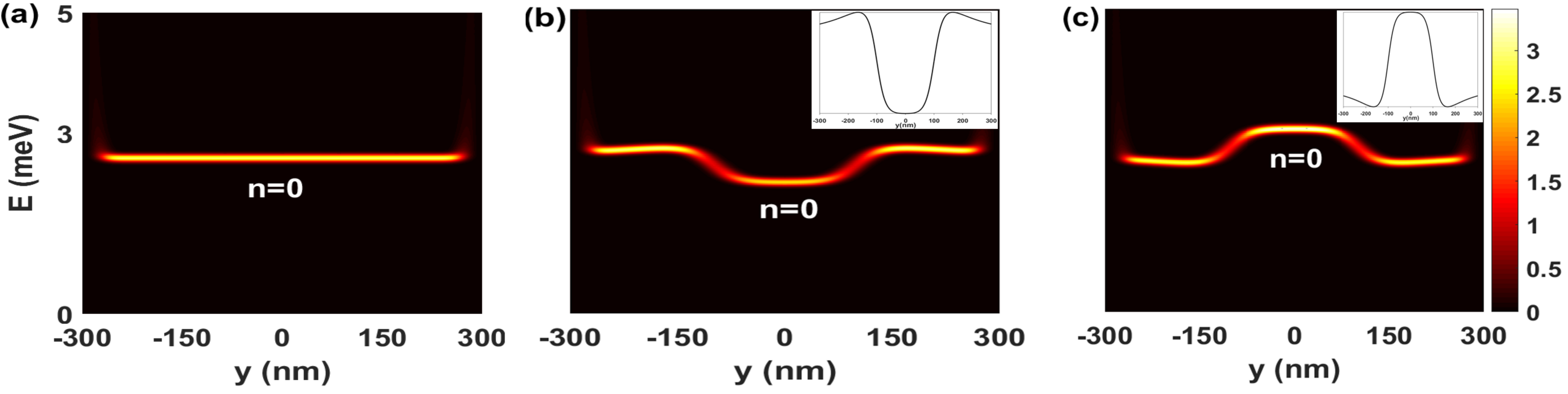}}
\caption{{\it (color online)} LDOS of lowest energy energy level at $ B_a $= 3 T and $ \Gamma $=0.06 meV for (a) zero magnetic modulation, (b) magnetic split and (c) magnetic strip gate. LDOS replicates the magnetic field profile for non zero modulation whereas does not show any variation for zero modulation. }
\label{fig:LDOSvs_magneticgate}
\end{figure*}
The LDOS of $n=0$ level is shown in Fig. \ref{fig:LDOSvs_magneticgate} for $B_a$ =3 T and $\Gamma $=0.06 meV. LDOS does not show variation when the magnetic modulation is absent (Fig. \ref{fig:LDOSvs_magneticgate} (a)). The energy levels are degenerate with respect to $ k_{x} $ (or the location of the centre of oscillator) for zero magnetic modulation. The modulated magnetic field lifts Landau level degeneracy near the center of the channel. This leads to the formation of states of lower energy at the center in the split gate (Fig. \ref{fig:LDOSvs_magneticgate}(b)) and of higher energy in the strip gate (Fig. \ref{fig:LDOSvs_magneticgate}(c)).
\begin{figure*}
\subfloat{\includegraphics[width=2\columnwidth,height=1
\columnwidth]{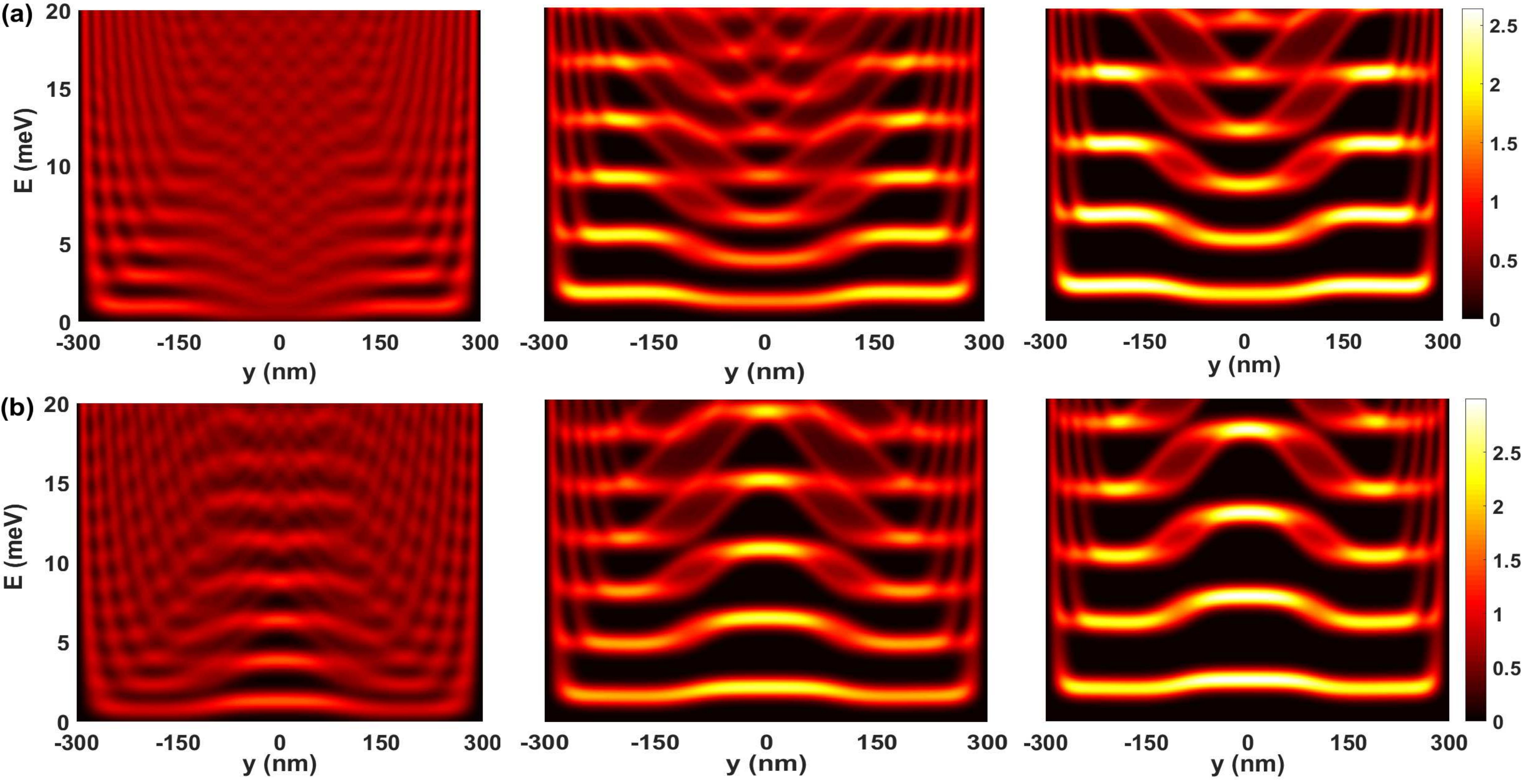}}
\caption{{\it (color online)} LDOS for a fixed channel width $ W_{e} $= 600 nm and $ \Gamma $=0.6 meV  at different values of $ B_{a} $=1 T, 2 T and 2.5 T for the split (a) and strip (b) gate. At $ B_{a} $=2.5 T, the magnetic minibands are separated from each other in the strip gate while they overlap in the split gate.}
\label{fig:LDOSvs_magneticfield}
\end{figure*} 

At high magnetic field ( $ \omega_{a} >> \omega^{\upharpoonleft \! \upharpoonright/\downharpoonright \! \upharpoonright} $), the quantized energy levels of the magnetic edge states (about the center of the channel) can be described using an approximate energy relation $\bar{E}_{n}(\bar{k}_{x})=(n+1/2)\hbar(\omega^{\upharpoonleft \! \upharpoonright/\downharpoonright \! \upharpoonright}_{m}(\bar{k}_{x})+\omega_{a})$ where $\omega^{\upharpoonleft \! \upharpoonright/\downharpoonright \! \upharpoonright}_{m}(\bar{k}_{x})=\frac{e B^{\upharpoonleft \! \upharpoonright/\downharpoonright \! \upharpoonright}_{m}(\bar{k}_{x})}{m^{\ast}}$ and $ \omega_{a}= \frac{e B_{a}}{m^{\ast}} $. The group velocity ($ dE_{n,\bar{k}_{x}}/d\bar{k}_{x} $) at y=$W_{m}/2$ is positive for magnetic split gate while it is negative for magnetic strip gate. The chirality of magnetic edge states reverses in the magnetic strip gate. As a result, the LDOS is a convenient tool to visualise the modification of energy levels by the combined effects of the inhomogeneous magnetic field and the electrostatic potential which is experimentally accessible to scanning tunneling microscopy \cite{Hamers}.
\begin{figure*}
\subfloat{\includegraphics[width=2\columnwidth,height=1
\columnwidth]{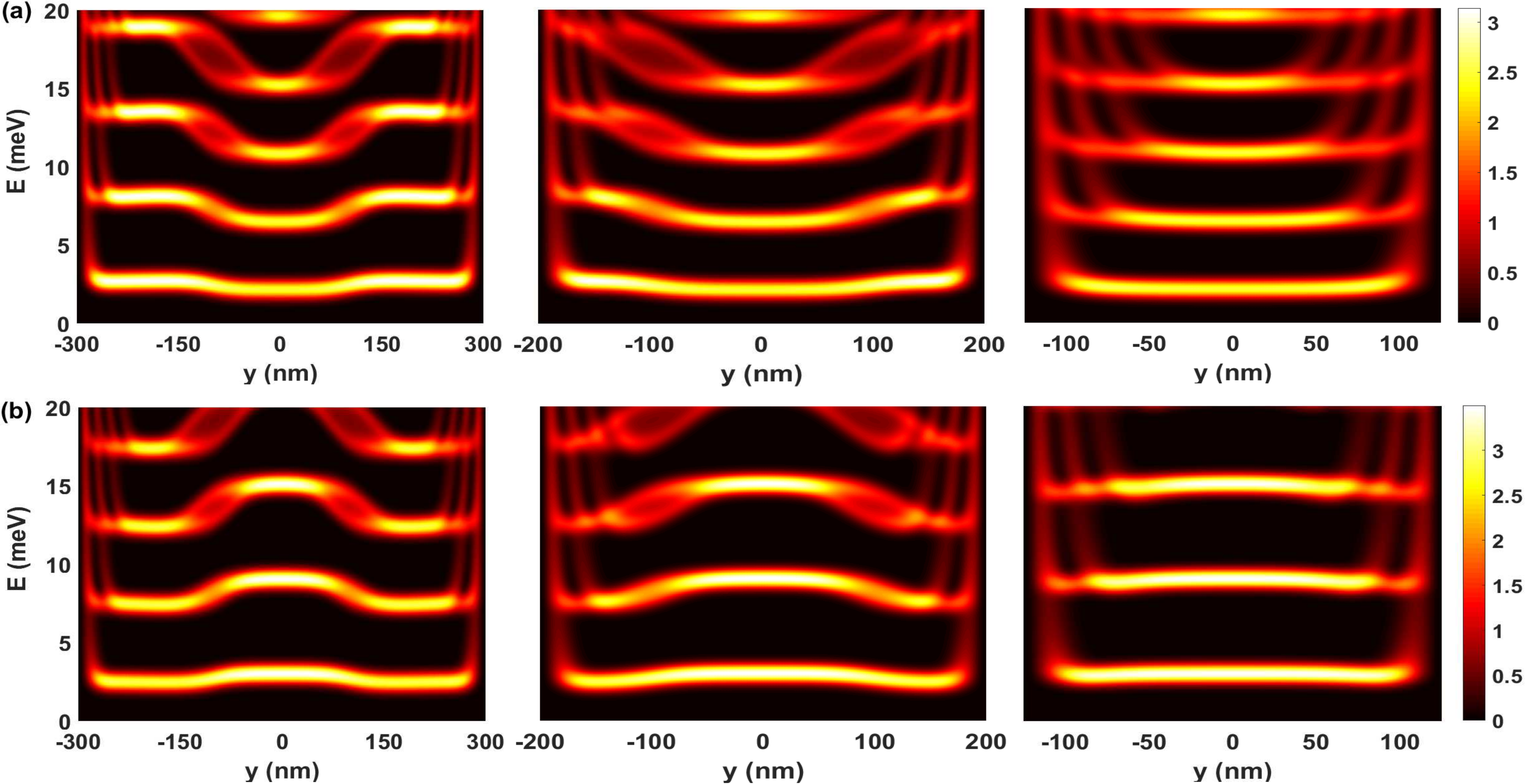}}
\caption{{\it (color online)} LDOS for a fixed $ B_a $= 3 T and $ \Gamma $=0.6 meV at channel widths $ W_e $=600 nm, 400 nm and 250 nm in the split (a) and strip (b) gate. As channel width decreases, overlap of magnetic and electrostatic edge states increases resulting in increasing back scattering in the strip gate (b).}
\label{fig:LDOSvs_electrostatic_width}
\end{figure*}      

Fig. \ref{fig:LDOSvs_magneticfield} shows the LDOS computed at different values of the applied magnetic field $B_a$=1 T, 2 T and 2.5 T in the split and strip gate channels of constant width $ W_e $= 600 nm. In this figure, the formation of a subset in the magnetic minibands in the central region is appearing due to the interferences between magnetic edge states in different Landau levels.     
The nature of these patterns can be understood when LDOS is plotted for a weaker impurity broadening (Supplementary Fig. 1 in \cite{Supple}). 
Since the LDOS is directly proportional to the probability density of the electrons (Eq. \ref{LDOS-1}), the number of branches in the dispersion curves demonstrates the number of nodes in the wavefunction of the magnetic edge states. The finite gradient of the energy spectrum therefore gets split into $n+1$ branches for the $n$-th Landau level where the magnetic field has finite gradient (Fig. 1 in supplemenatry information \cite{Supple} for $ \Gamma $=0.1 meV).   

However, the splitting vanishes where the magnetic field gradient approximately zero and the dispersion is flat as at $y  $=0 nm and $y  $=200 nm for the profiles considered in this paper. Also the energy separation between the adjacent Landau bands varies over the region and become large when magnetic field gradient is steeper. The splitting as well as the overlap between the magnetic edge states corresponding to different Landau bands is naturally smeared for larger impurity broadening (Fig. \ref{fig:LDOSvs_magneticfield} (a) and (b) for $ B_{a}$=2 T and 2.5 T). At lower magnetic field, the relative effect of  the magnetic modulation is much stronger and splitting is barely observed as the magnetic edge states overlap with each other (Fig. \ref{fig:LDOSvs_magneticfield} (a) and (b) for $ B_{a}$=1 T). 
         
Magnetic minibands overlap in energy at low magnetic field.    However above a critical magnetic field, energy gaps open between magnetic minibands ($ B_a $=2.5 T in Fig. \ref{fig:LDOSvs_magneticfield} (b)). Gap opens in the magnetic minibands when the energy gap between consecutive energy levels is larger than the magnetic band width. The magnetic band width ($ \Delta_{n} $) is of the form $ \Delta_{n} =(n+1/2)\frac{\hbar e B_m^{peak}}{m^{\ast}}$. In the split gate, negative magnetic modulation at the center decreases the energy gap between consecutive energy levels by $  \frac{\hbar e}{m^{\ast}} B_m^{peak}  $ amount from $ \frac{\hbar e}{m^{\ast}} B_a$.          
Therefore, the applied magnetic field at which gaps open in the magnetic minibands in the split gate which is the difference between the $ n^{th} $ Landau level in the unmodulated region and the $ (n-1)^{th} $ Landau level in the modulated region
 follows the relation 
\beq
\frac{\hbar e}{m}(B_a-B_m^{peak})\approx \Delta_{n}
\label{eq:magnetic_field_fermienergy2}
\eeq  
Now by inserting   \ $ (n+1/2)= \frac{B_{F}}{B_{a}}$ in Eq. \ref{eq:magnetic_field_fermienergy2} we get 
\bea
& & B_a^{2}+B_m^{peak}B_a-B_F B_m^{peak}=0 
\eea  
where $B_F=\hbar k_{F}^{2}/2e$ and $ k_{F}  $ is the Fermi wave vector. The applied magnetic field at which minigaps open in the split gate is of the form
\bea
B_{a}^{\upharpoonleft \! \upharpoonright} &=& \frac{B_m^{peak}+\sqrt{(B_{m}^{peak})^{2}+4B_FB_m^{peak}}}{2}
\label{eq:gapopen_magneticfield}
\eea
We obtain corresponding magnetic field for the strip gate by replacing $ B_m^{peak} $ with $- B_m^{peak} $ in Eq. \ref{eq:gapopen_magneticfield}.
We evaluate $B_{a}^{\upharpoonleft \! \upharpoonright}$=2.8 T for split gate and $B_{a}^{\downharpoonright \! \upharpoonright}$=2.2 T for the strip gate (Fig. \ref{fig:LDOSvs_magneticfield} (b)) using $B_{m}^{peak}=$0.65 T. The gap opening in the magnetic minibands leads to large amplitude oscillation in the density of states (DOS) and also in the conductivity.    
   
Fig. \ref{fig:LDOSvs_electrostatic_width} plots the LDOS at a fixed value of $ B_a $= 3 T for decreasing values of channel width $ W_e $=600 nm, 400 nm and 250 nm. For wider channel width ($ W_e $=600 nm), the magnetic and electrostatic edge states are separated by $\sim 8 \ell_{b}$  resulting in small overlap of their wavefunction. But as electrostatic confinement increases, the overlap of the electronic wavefunction increases.  When $ W_e <W_m $, the magnetic edge states are depleted leaving behind only the electrostatic edge states. Thus, as one decreases $ W_e$, the electrostatic edge states cross over magnetic edge states.  

\begin{figure}[h!]
\subfloat{\includegraphics[width=1\columnwidth,height=1.4
\columnwidth]{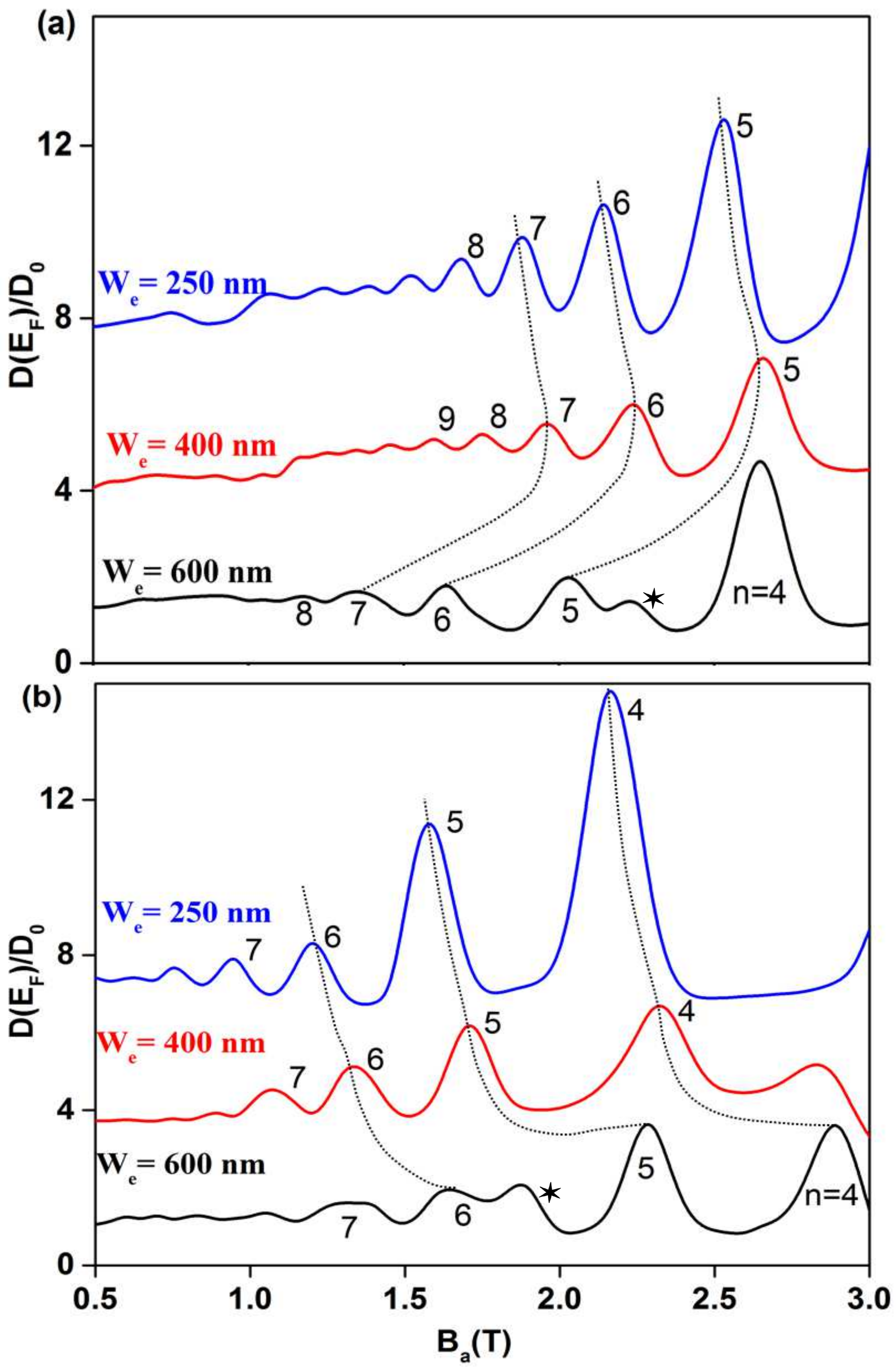}}
\caption{{\it (color online)} Density of states at the Fermi energy vs applied magnetic field in the split gate (a). Peaks shift to the higher magnetic field for decreasing the electrostatic confinement width from 600 nm to 400 nm while peaks shift to lower magnetic field in the split gate (b). $ \ast $ symbols denote the small amplitude oscillations due to magnetic minibands. }
\label{fig:DOSvsBa_electrostatic_width}
\end{figure}

We have plotted the DOS of both devices as a function of applied magnetic field in Fig. \ref{fig:DOSvsBa_electrostatic_width}. The DOS is given as   
\beq
 D(E)= D_{0}\hbar \omega \sum_{n}\int dk_{x}P_{imp}(E-E_{n,k_{x}})
 \label{eq:DOS_net_broadening}
\eeq
Where Eq. \ref{eq:DOS_net_broadening} accounts for both both modulation and impurity broadening to the energy levels.  
Small amplitude oscillations at lower $ B_{a} $ corresponds to the overlap of magnetic minibands. At high $ B_{a} $, the small amplitude oscillations also occur besides the high amplitude oscillation for wider channel width $ W_{e} $=600 nm ($ \ast $ symbols in Fig. \ref{fig:DOSvsBa_electrostatic_width} ). These oscillations appear due to the existence of magnetic minibands. But as $ W_{e} $ decreases, the depletion of magnetic minibnads (as shown in Fig. \ref{fig:LDOSvs_electrostatic_width} for $ W_{e} $=400 nm and 250 nm) causes the small amplitude oscillation to die out. Also, the peaks shift to higher magnetic field (split gate in Fig. \ref{fig:DOSvsBa_electrostatic_width} (a)) or lower magnetic field (strip gate in Fig. \ref{fig:DOSvsBa_electrostatic_width} (b)) as $ W_{e} $ decreases from 600 nm to 400 nm. 

The shifting of peaks position as a function of $ W_e $ can be understood from the LDOS plot for various values of $ W_{e} $ which is shown in Fig. \ref{fig:LDOSvs_electrostatic_width}. The magnetic minibands are successively depleted as we increase the electrostatic confinement. When $ W_{e} $ decreases from 600 nm to 400 nm, the peak in the DOS shift from the edge to the band center which is lower in energy (Fig. \ref{fig:LDOSvs_electrostatic_width} (a)). Therefore, decreasing $ W_{e} $ causes the maxima of DOS to shift lower in energy. To keep the highest occupied band aligned with the Fermi level, a higher magnetic field is required resulting in shifting the peaks position to the higher magnetic field (Fig. \ref{fig:DOSvsBa_electrostatic_width} (a)). As $ W_{e} $ further decreases, peaks shift to lower magnetic field due to the decreasing value of the Fermi energy. 

In contrast, the DOS peaks in the strip gate shift to the lower magnetic field when $ W_{e} $ decreases from 600 nm to 400 nm. In this case, positive modulation at the center lowers the energy of magnetic minibands relative to the center (Fig. \ref{fig:LDOSvs_electrostatic_width} (b)). This results in DOS maxima to shift higher in energy. Thus, less applied magnetic field is needed to keep the central Landau levels aligned with the Fermi level causing LDOS peaks to shift to lower magnetic field. From $ W_{e} $=400 nm to 250 nm, peaks shift to the lower magnetic field because of the decrease in electron density (or Fermi energy).  
\section{Conductivity tensor and Magnetoresistance}
\begin{figure*}
\subfloat{\includegraphics[width=2\columnwidth,height=0.8
\columnwidth]{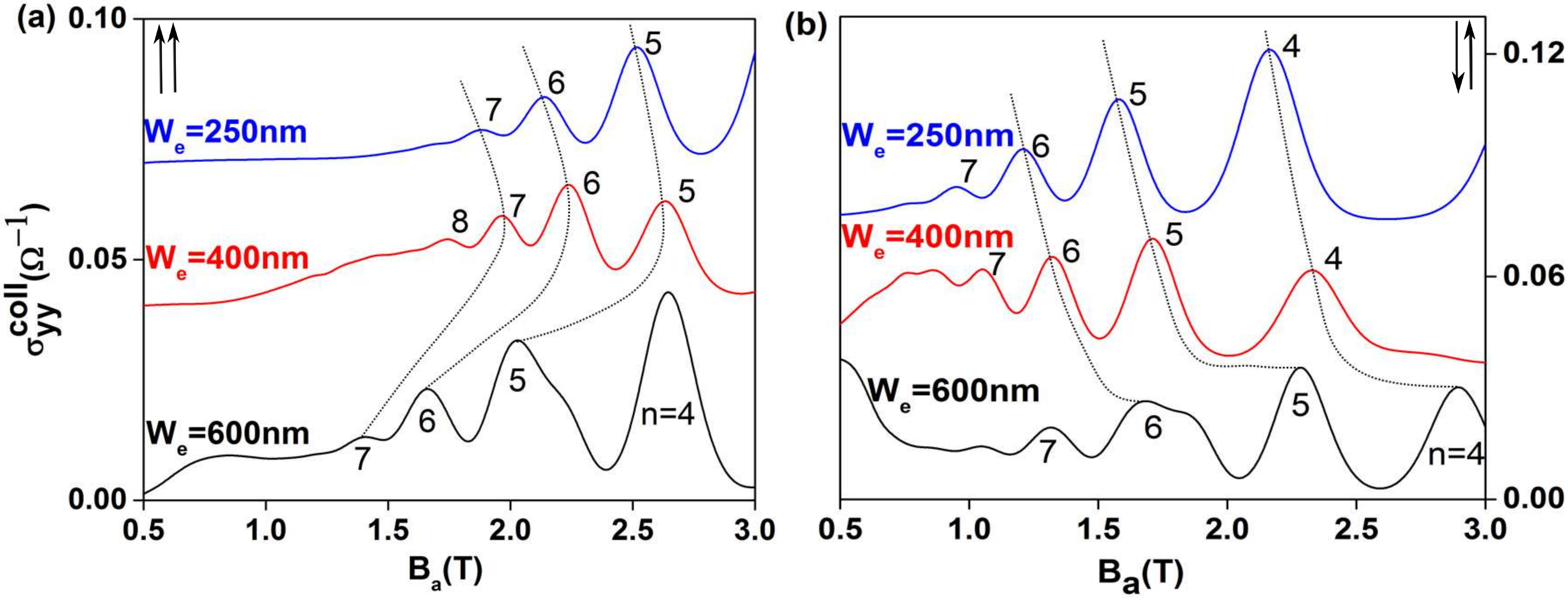}}
\caption{{\it (color online)} Collision conductivity as a function of applied magnetic field in the split (a) and strip (b) gate. Arrows indicate parallel/anti-parallel motion of edge states. Back scattering in the strip gate enhances the peaks amplitude as channel width decreases.  }
\label{fig:collision_conduc}
\end{figure*}
 \begin{figure*}
\subfloat{\includegraphics[width=2\columnwidth,height=0.8
\columnwidth]{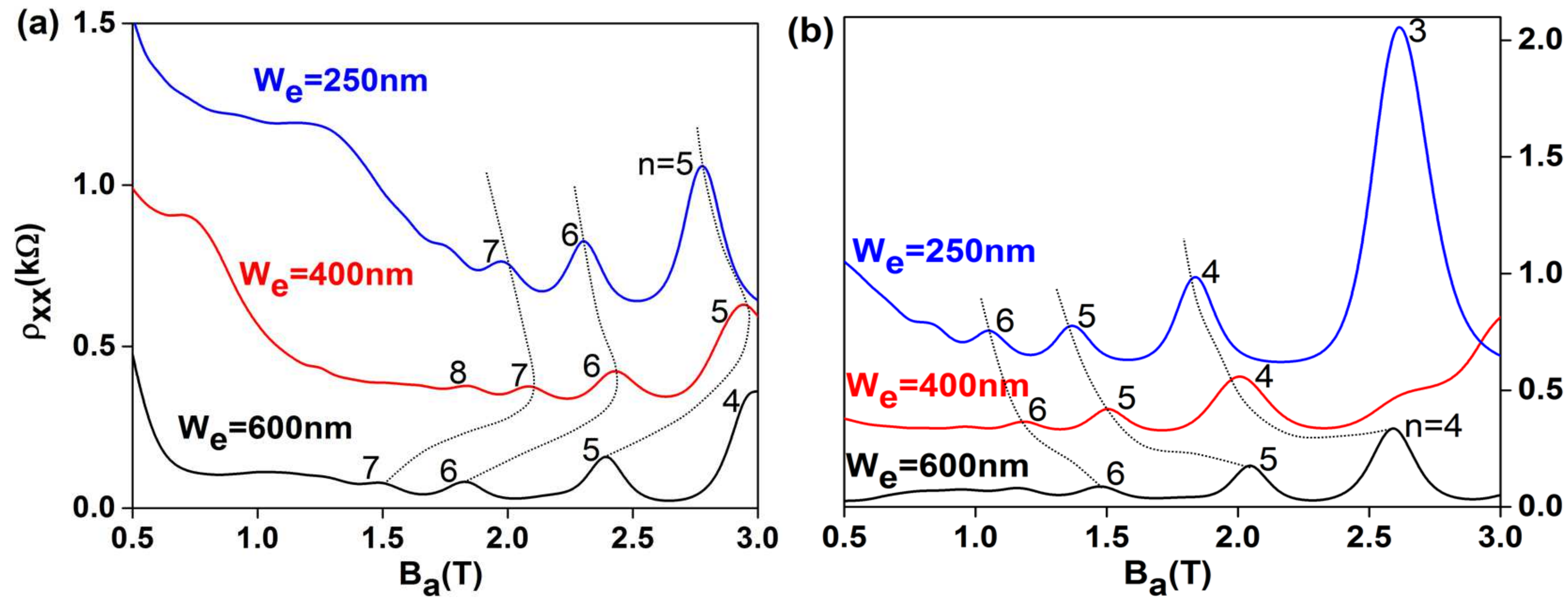}}
\caption{{\it (color online)} Magneto-resistance for decreasing channel width in the split (a) and strip (b) gate.}
\label{fig:resistivity}
\end{figure*}
We calculate the magnetoresistance of the strip and split gate systems to quantify the effects of different electronic structures and the chiral/non-chiral nature of edge state transport. We use linear response theory to determine the current density J as a response to an weak applied electric field E \cite{Abe}. In the presence of weak field and disorder, the conductivity tensor contains both diagonal and non-diagonal parts which come from the diagonal/non-diagonal part of the current density operator, has been calculated in \cite{Charbonneau}. The conductivity tensor is
\bea
\sigma_{\mu,\nu}(\omega)= \sigma_{\mu,\nu}^{d}(\omega)+\sigma_{\mu,\nu}^{nd}(\omega) ~~\mu,\nu=x,y,z
\eea
Where $ d $ and $ nd $ are the diagonal and non-diagonal parts of the conductivity tensor. We calculate various contributions to the conductivity tensor in the static limit ($\omega \rightarrow 0$). Diagonal components of conductivity tensor consist of the band conductivity and the scattering/collisional conductivity. 
\bea
\sigma^{d}_{\mu,\nu}(\omega)= \sigma_{\mu,\nu}^{band}(\omega)+\sigma_{\mu,\nu}^{coll}(\omega) 
\eea

The band contribution of the conductivity is of the form
\bea
\sigma_{\mu \nu}^{band}&=&\frac{\beta e^{2}}{A}\sum_{n,k_{x}}\int dE ~P_{imp}(E-E_{n,k_{x}})f_{E}(1-f_{E})\tau(E)\nn\\
&\times&\upsilon_{\mu}^{n,k_{x}}~\upsilon_{\nu}^{n,k_{x}}
\label{band-cond}
\eea
where $\beta=1/k_B T$, A is the area of the sample, $f_{E}$ is the Fermi-Dirac distribution function, $\tau(E)$ is the relaxation time and $\upsilon_{\mu}^{n,k_{x}}$ is the velocity operator given as $\frac{1}{\hbar}\frac{\partial E_{n,k_{x}}}{\partial k_{\mu}}$. Electrons are free particles along the $ x $ direction while localized along the $ y $ direction {\it i.e} $\upsilon_{y}=0$. Therefore, band conductivity along the $ y $ direction is zero {\it i.e} $\sigma_{yy}^{band}=0$ and also $\sigma_{xy}^{band}=\sigma_{yx}^{band}= 0$. The band conductivity along the $ x $ direction is 
\bea
\sigma_{xx}^{band} &=&  \frac{e^{2}}{h}\frac{\tau}{L_{y}\hbar\sqrt{\pi}\Gamma}\sum_{n}\int dE ~\left( -\frac{\partial f}{\partial E}\right) \int dk_{x}\nonumber\\
& \times&\exp{\left( -\frac{(E-E_{n,k_{x}})^{2}}{\Gamma^{2}}\right) }\bigg{\vert} \frac{\partial E_{n,k_{x}}}{\partial k_{x}}\bigg{\vert}^{2}
\eea

Collision conductivity considers the transport through localized states in the presence of impurities. We consider $\sigma_{xx}^{col}=\sigma_{yy}^{col}$ due to isotropic impurity scattering. 
Collisional contribution to the conductivity ($\sigma_{xx}^{col}$) with broadening is given by \cite{Shi} 
\bea
\sigma_{xx}^{col} &=& \frac{\beta e^{2}}{A}\sum_{\xi \xi^{'}} \int dE \int dE^{'} P(E-E_{\xi})P(E^{'}-E_{\xi^{'}})\nn\\
&& f(E)[1-f(E^{'})]W_{\xi \xi^{'}}(E,E^{'}) (\alpha_{x}^{\xi}-\alpha_{x}^{\xi^{'}})^{2} 
\label{coll-cond}
\eea
where $W_{\xi \xi^{'}}$ is the transition rate between $\vert \xi \rangle$ and $\vert \xi^{'} \rangle$  and $\alpha_{x}^{\xi}=\langle nk_{x} \vert x \vert nk_{x} \rangle$ is the expectation value of position operator and $\vert \xi \rangle= \vert n,k_{x} \rangle$ . We consider elastic scattering for which $f(E)=f(E^{'})$. The transition rate is given as
\bea
W_{\xi \xi^{'}}(E,E^{'}) &=& \frac{2 \pi N_{I}}{A \hbar} \sum_{q} \vert U_{q} \vert^{2}\vert \langle \xi \vert e^{iq.r}\vert \xi^{'} \rangle \vert^{2} \delta (E -E^{'})\nn\\
\label{tran-rate}
\eea
$U_{q}$ is the Fourier transform of the impurity potential
$ U(r-R)=\frac{e^{2}}{4 \pi \epsilon \epsilon_{0} \vert r-R \vert }e^{-k_{s}\vert r-R \vert}  $; r and R are the position of the electron and impurity, $k_s$ is the screening wave vector and $N_I$ is the impurity density. We consider only the dominant term $n=n^{'}$ (intra level scattering) and next-nearest inter level scattering term {\it i.e} $n-n^{'}=\pm 1$ which is sufficient because scattering rate decreases exponentially with the distance between centres of oscillator. $k_{s}=q_{s}k_{F}$, $q_{s}=\frac{2^{3/2}m e^{2}}{\epsilon \hbar^{2}\sqrt{4\pi n_{s}}} \rightarrow k_{s} \simeq k_{F}$, $U_{0}=\frac{e^{2}}{2\epsilon\epsilon_{0}}$ and $l_{b}=\sqrt{\hbar/eB}$. 

The collisional conductivity is shown in Fig. \ref{fig:collision_conduc}. At a wider channel width ($W_e$ =600 nm), the amplitude of peaks in the collisional conductivity ({\it e.g.} $ n=6 $) are larger in the strip gate than the split gate.  As the channel width decreases from 600 nm, the overlap of magnetic and electrostatic edge states increases (as shown in Fig. \ref{fig:LDOSvs_electrostatic_width}) resulting in increasing scattering between magnetic and electrostatic edge states. The increase on $ \sigma_{yy} $ peaks  is more pronounced in the strip gate where the edge states drift anti parallel due to back scattering between the pair of edge states. As a result, the peak amplitude increases as compared to the split gate where the intra edge scattering does not effect the conductivity since the magnetic and electrostatic edge states drift in the same direction. Hence, the collisional conductivity reflects the chiral/non-chiral nature of the edge states of the two devices. 

We have calculated resistivity from the band conductivity (Eq. \ref{band-cond}), collision conductivity (Eq. \ref{coll-cond}) and the non-diagonal conductivity (Eq. II.2 in \cite{Supple}) which we combined in the plot of the longitudinal resistivity shown in Fig. \ref{fig:resistivity}. The effect of reducing channel width is to shift resistance peaks to lower magnetic field in the split gate system where central magnetic field is lower.  In contrast resistance peaks move to higher magnetic field in the strip gate according to the same mechanism as in Fig. \ref{fig:DOSvsBa_electrostatic_width}. The amplitude of the peaks increases when $ W_{e} $ decreases from 600 nm to 400 nm in the magnetic strip gate because of the increasing overlap of edge states (as $ n=5 $ in Fig. \ref{fig:collision_conduc} (b)). Note that the higher amplitude of resistance peaks in the case of the strip gate relative to the split gate is consistent with the existence of backscattering between counter-propagating edge states in the strip gate.
\section{Conclusions}
In conclusion, we have studied the properties of a 2DEG exposed to two types of modulated magnetic field profile. We have calculated LDOS which shows the difference in electronic structure of  the split and strip gate devices. LDOS demonstrates the formation of magnetic edge states near the center of the gate and electrostatic edge states at the boundary. Also, LDOS shows interference originating magnetic minibands subset at the center of the gate. In the split gate, magnetic and electrostatic edge states drift parallel. In contrast, the opposite magnetic field gradient in the strip gate changes the drifting direction of magnetic edge states resulting anti-parallel motion of magnetic and electrostatic edge states. 

The calculated collisional conductivity depends on the relative drift directions of  magnetic and electrostatic edge states. This dependence is borne by the shape of the wavefunction in the scattering matrix element. While the diffusion and non-diagonal conductivities are independent of drift direction and only contribute to the diffusive conduction regime at low magnetic field.  
The resistance peaks at high magnetic field are dominated by the collisional conductivity and larger in the case of strip gate than the split gate device as channel width decreases.  This indicates stronger backscattering in the case of edge states which propagate in opposite directions (strip gate) than in the same direction (split gate).  This suggest that the chiral/non-chiral nature of the bulk and edge conducting states might be probed through quantum transport measurements. 

This work was supported by a UGC-UKIERI 2013-14/081 thematic award. P. Mondal is supported by UGC fellowship. 

\def \bb{\bibitem}

\end{document}